\documentclass[12pt]{article}
\usepackage{amsmath,amsfonts,epsf}

\renewcommand{\Re}{\operatorname{Re}}
\renewcommand{\Im}{\operatorname{Im}}
\newcommand{\Tr}{\operatorname{Tr}}

\newcommand{\A}{\boldsymbol A}
\renewcommand{\r}{\boldsymbol r}
\newcommand{\ro}{\boldsymbol r_0}
\newcommand{\erfc}{\operatorname{erfc}}
\newcommand{\sign}{\operatorname{sign}}
\newcommand{\ds}{\displaystyle}

\begin{document}

\noindent
mpi-pks/9903009
\vspace{3.0cm}

\centerline{\LARGE Semiclassical Treatment of Diffraction}
\vspace{0.3cm}
\centerline{\LARGE in Billiard Systems with a Flux Line}
\vspace{2.0cm}
\centerline{\large Martin Sieber}
\vspace{0.5cm}

\centerline{
Max-Planck-Institut f\"ur Physik komplexer Systeme,
N\"othnitzer Str.\ 38, 01187 Dresden, Germany
\footnotemark}
\footnotetext{present address}
\vspace{0.1cm}
\centerline{
Abteilung Theoretische Physik, Universit\"at Ulm, 89069 Ulm, Germany}

\vspace{5.0cm}
\centerline{\bf Abstract}
\vspace{0.5cm}

In billiard systems with a flux line semiclassical
approximations for the density of states contain
contributions from periodic orbits as well as from
diffractive orbits that are scattered on the flux line.
We derive a semiclassical approximation for diffractive
orbits that are scattered once on a flux line. This
approximation is uniformly valid for all scattering angles.
The diffractive contributions are necessary in order that
semiclassical approximations are continuous if the position
of the flux line is changed.

\vspace{2.5cm}

\noindent PACS numbers: \\
\noindent 03.65.Sq ~ Semiclassical theories and applications. \\
\noindent 05.45.-a ~ Nonlinear dynamics and nonlinear dynamical systems. \\
\noindent 05.45.Mt ~ Semiclassical chaos (``quantum chaos'').

\newpage

\section{Introduction}
\label{secintro}

Billiard systems with flux lines are simple dynamical
systems that can model the typical behaviour of low
dimensional quantum systems. One main application
has been in the investigation of universal properties
of quantum systems that are chaotic in the classical limit 
\cite{BR86}.
The high lying states of these systems have statistical
properties that depend only on the symmetries of the system
and can be described by random matrix theory
\cite{Boh91,Haa92,Meh91}. Ordinary billiards are
used to model systems that are invariant under
time inversion and are described by the Gaussian orthogonal
ensemble (GOE). The introduction of an additional flux line
allows to study a further universality class since the flux
line breaks the time-reversal symmetry. If the flux strength
is sufficiently large (when considering levels in a fixed
energy range) the corresponding universality class is
that of the Gaussian unitary ensemble (GUE), but by
varying the flux strength one can investigate also the
GOE-GUE transition or parametric correlations
in the GUE regime. Another application is for
integrable billiard systems where the introduction
of a flux line can break the integrability of the
quantum billiard \cite{DJM95}, or to use billiards
with a flux line to model the properties of quantum dots
(see e.g.\ \cite{BLM96}).

A major advantage in all these applications is that
the classical trajectories are not changed by the
presence of a flux line, except for the set of measure
zero that hit the flux line. In a semiclassical
analysis the same set of periodic orbits appears
independent of the flux strength and only the
semiclassical contributions of these orbits are
changed. For example, in approximations to the
density of states the periodic orbit contributions
have an additional phase $2 \pi m \alpha$ where
$\alpha$ is related to the flux strength and $m$
is the number of times the orbit winds around the
flux line. However, this is not the only semiclassical
effect of a flux line. Its presence leads also to wave
diffraction that can be semiclassically described
by an additional set of trajectories that are
closed but not periodic and which start and end on
the flux line. In semiclassical arguments these so-called
diffractive orbits are often neglected since their
semiclassical contribution is estimated to be of
order $\sqrt{\hbar}$ smaller than the contributions
of periodic orbits. Such an estimate is, however, only
valid if the scattering angle is not too close to
the forward direction. In the forward direction
diffractive orbits contribute in the same order
as periodic orbits.

In this article we investigate the semiclassical
contributions of diffractive orbits to the density
of states. We derive an approximation for orbits
that are scattered once on a flux line. This 
approximation is valid for all scattering angles,
also in the forward scattering direction.
We show that these contributions are necessary in
order to cancel discontinuities in periodic orbit 
contributions that occur when the position of 
the flux line is changed and crosses a periodic orbit.
We also discuss the importance of diffractive
orbits in the semiclassical limit.

The article is organised as follows.
In section 2 we consider the scattering on a flux line
in a plane and derive a uniform approximation for 
the Green function. With this input we derive in section 3
a uniform approximation for semiclassical
contributions to the density of states from isolated
diffractive orbits that are scattered once on a 
flux line. As an example of a system with non-isolated
diffractive orbits we consider in section 4 the
integrable circular billiard with a flux line in
its center. The results are discussed in section 5.

\section{A flux line in a plane}
\label{secplane}

The scattering of a wave function on a flux line
in a plane without boundaries has been studied in
great detail since the problem was first treated
by Aharonov and Bohm \cite{AB59}. There exists,
for example, an exact integral representation
for the propagator. A review on the quantum
effects of electromagnetic fluxes is given in
\cite{OI85}. In this section we review some of
the results and derive a uniform
approximation for the Green function that will
be used in the following section.

The Schr\"odinger equation for a particle with mass $M$ and
charge $q$ in a magnetic field in two dimensions is given in
Gaussian units by
\begin{equation} \label{diffpla1}
\frac{1}{2 M} \left[ \frac{\hbar}{i} \nabla - \frac{q}{c} 
\A \right]^2 \Psi(\r) = E \Psi(\r) \; ,
\end{equation}
and for a flux line the vector potential can be chosen as
\begin{equation} \label{vecpot}
\A = \frac{\Phi}{2 \pi r} \hat{\phi} \; , \qquad
\nabla \times \A = \Phi \delta(\r) \hat{z} \; , \qquad 
\nabla \cdot \A = 0 \; ,
\end{equation}
which describes a magnetic field with flux $\Phi$ that
is concentrated in form of a delta-function in the origin.

In polar coordinates the Schr\"odinger equation with
vector potential (\ref{vecpot}) has the form
\begin{equation} \label{diffpla2}
- \left[ \frac{\partial^2}{\partial r^2} 
+ \frac{1}{r} \frac{\partial}{\partial r} + \frac{1}{r^2}
\left( \frac{\partial}{\partial \phi} - i \alpha \right)^2
\right] \Psi(r, \phi) = k^2 \Psi(r,\phi) \; ,
\end{equation}
where $\alpha = q \Phi/(2 \pi \hbar c)$ and $k = \sqrt{2 M E}/\hbar$.
The solutions that are regular in the origin are given by
\begin{equation} \label{solpla}
\Psi_{m,k}(r,\phi) = \sqrt{\frac{k}{2 \pi}}
\, J_{|m-\alpha|}(k r) \, \exp\{ i m \phi \} \; ,
\end{equation}
where the normalisation is chosen such that
\begin{equation}
\int_0^{2 \pi} \! d \phi \; \int_0^\infty \! dr \; r \, 
\Psi_{m,k}(r,\phi) \, \Psi_{m',k'}^*(r,\phi)
= \delta_{m,m'} \, \delta(k - k') \; ,
\end{equation}
as follows from the orthogonality relations of the exponential and
the Bessel functions \cite{Wat66}. 
It is sufficient to consider only values $0 \leq \alpha
\leq 0.5$ since the solutions for other values of $\alpha$
can be obtained by a multiplication of a ($\phi$-dependent)
phase factor and possibly a complex conjugation.

With the normalised solutions (\ref{solpla}) the propagator
can be written down directly
\begin{align} \label{propa}
K_\alpha(\r,\ro,t) = \, & \sum_{m=-\infty}^\infty \int_0^\infty \! dk \;
\Psi_{m,k}(r,\phi) \, \Psi_{m,k}^*(r_0,\phi_0) \,
\exp\left\{-\frac{i}{\hbar} \frac{\hbar^2 k^2}{2 M} t \right\}
\notag \\ &
= \sum_{m=-\infty}^\infty \frac{M}{2 \pi i \hbar t} \,
J_{|m-\alpha|} \left( \frac{r r_0 M}{\hbar t} \right) \,
\exp \left\{ i m (\phi - \phi_0) - \frac{M(r^2 + r_0^2)}{2 i \hbar t}
- i \frac{\pi}{2} |m- \alpha| \right\} \; ,
\end{align}
and in the same way one obtains an exact representation for the Green
function
\begin{align} \label{greenex}
G_\alpha(\r,\ro,E) = \, & \lim_{\varepsilon \rightarrow 0} \frac{1}{i \hbar}
\int_0^\infty \! dt \; K_\alpha(\r,\ro,t) \, \exp \left\{ \frac{i}{\hbar}
t (E + i \varepsilon) \right\} 
\notag \\ & 
= \lim_{\varepsilon \rightarrow 0} 
\sum_{m=-\infty}^\infty \int_0^\infty \! dk' \;
\Psi_{m,k'}(r,\phi) \, \Psi_{m,k'}^*(r_0,\phi_0) \,
\frac{1}{E + i \varepsilon - \frac{\hbar^2 {k'}^2}{2 M} }
\notag \\ &
= \sum_{m=-\infty}^\infty \frac{M}{2 i \hbar^2} \,
e^{i m (\phi - \phi_0)} 
J_{|m-\alpha|} (k r_<) \, H_{|m - \alpha|}^{(1)}(k r_>) \; ,
\end{align}
where $r_<$ and $r_>$ are the smaller and larger values
of $r$ and $r_0$, respectively.

For further evaluations the representation in the basis of
the angular momentum eigenstates is not convenient. One can
derive, however, from the sum in (\ref{propa}) an exact integral
representation for the propagator (see \cite{OI85}). It involves
an integral over Hankel functions. In the following we use
an approximation to this integral representation that 
is obtained after replacing the Hankel functions by their
leading asymptotic form. This gives a semiclassical approximation
for the propagator that is valid if $0 < \alpha < 1$ and
$M r r_0/t \gg \hbar$, i.\,e.\ for not too long times or
too close distances to the flux line.
\begin{align} \label{propuni}
K_\alpha(\r,\ro,t) \approx \, & 
\frac{M}{2 \pi i \hbar t} \exp \left\{ \frac{i M}{2 \hbar t} 
(\r - \ro)^2 + i \alpha (\phi - \phi_0) \right\}
\notag \\ &
- \frac{M \sin(\alpha \pi)}{\pi i \hbar t} \exp \left\{
\frac{iM}{2 \hbar t} (r + r_0)^2 + \frac{i}{2} (\phi - \phi_0) \right\}
\, K \left[ \sqrt{\frac{2 M r r_0}{\hbar t}} \, \cos \left( 
\frac{\phi - \phi_0}{2} \right) \right] \; .
\end{align}
The angles in (\ref{propuni}) have to be chosen such that 
$|\phi - \phi_0| \leq \pi$, and the function
$K(z)$ is described below. The semiclassical
propagator in (\ref{propuni}) consists of two parts.
The first part is almost identical
to the propagator in a free plane with the difference of
an additional phase proportional to $\alpha$. 
Semiclassically it can be interpreted as the contribution
from the direct path from $\r_0$ to $\r$.
The additional phase results from the dependence of the
Lagrangian on the vector
potential. The second term in (\ref{propuni}) describes the
scattering on the flux line and is discussed
in more detail in connection with the Green function.
Both terms are discontinuous in the forward direction
$|\phi - \phi_0| = \pi$ but the sum is continuous
and uniformly valid for all values of $\phi$ and
$\phi_0$.

\begin{figure}[thb] \label{figk}
\begin{center}
\mbox{\epsfxsize8cm\epsfbox{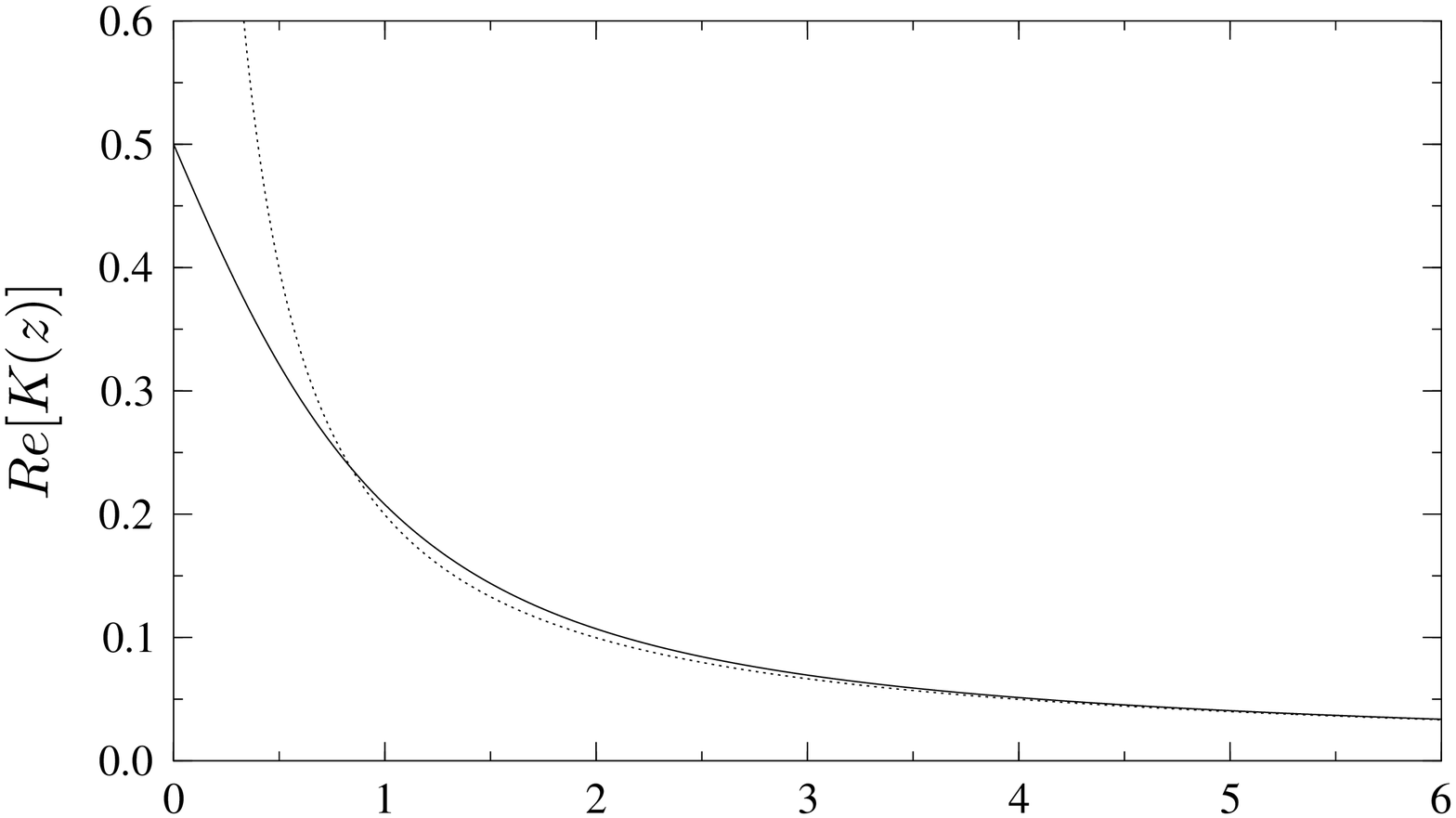}}
\mbox{\epsfxsize8cm\epsfbox{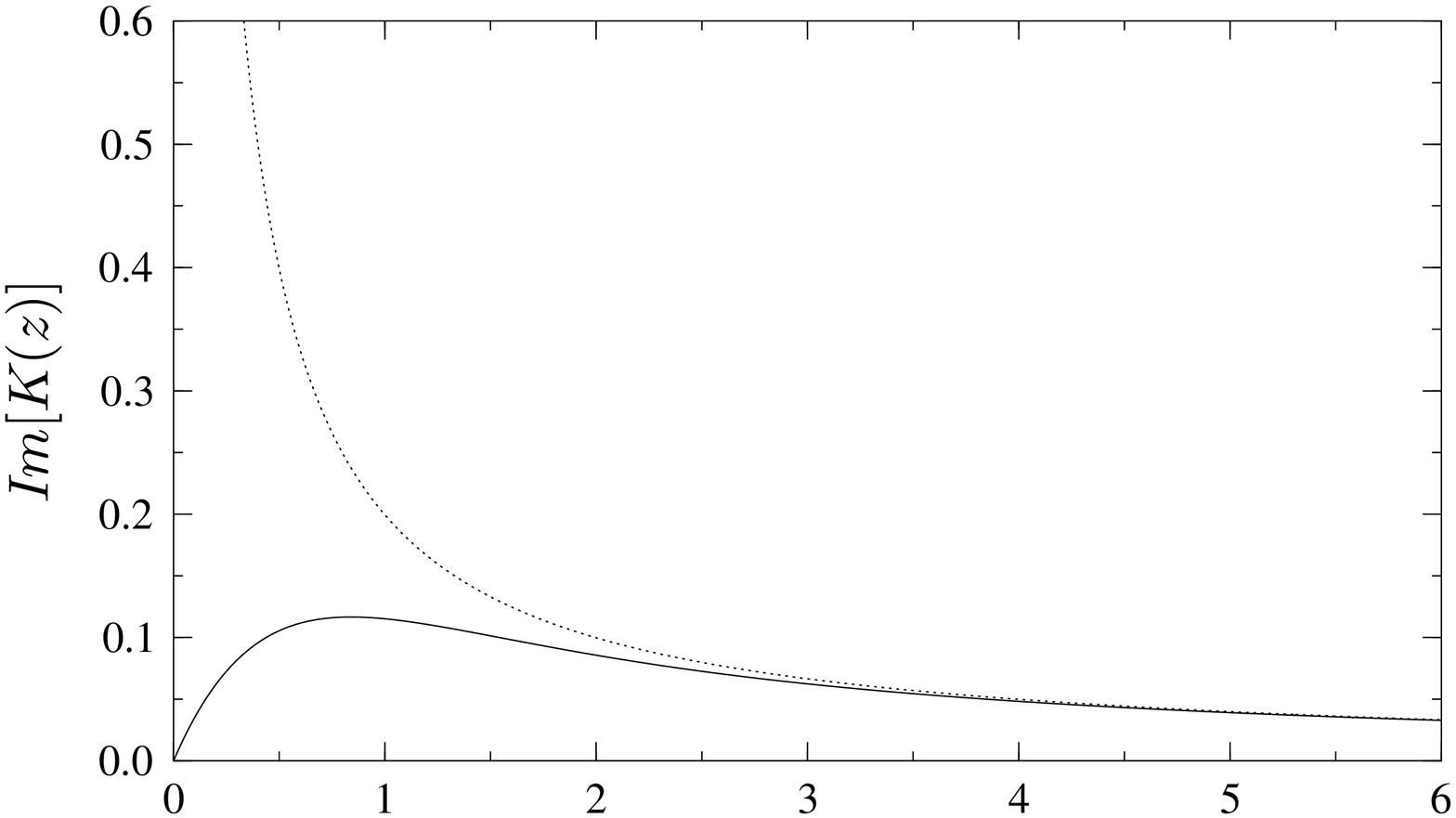}}
\end{center}
\caption{Real and imaginary parts of the
modified Fresnel function $K(z)$ (full line) and
its asymptotic approximation (dotted line).}
\end{figure}

The function $K(z)$ in (\ref{propuni}) is a modified Fresnel
function that is defined by
\begin{equation}
K(z) = \frac{1}{\sqrt{\pi}} \exp\left\{ - i z^2
- i \frac{\pi}{4} \right\} \int_z^\infty \! dy \; e^{i y^2}
= \frac{1}{2} e^{-i z^2} \erfc \left( e^{- i \pi/4} z \right) \; ,
\end{equation}
and it has the following limiting properties
\begin{equation} \label{kasym}
K(0) = \frac{1}{2} \; ; \qquad \qquad K(z) \sim \frac{e^{i \pi/4}
}{2 z \sqrt{\pi}} \; , \; \; |z| \rightarrow \infty \; , \; \;
-\frac{\pi}{4} < \arg(z) < \frac{3 \pi}{4} \; .
\end{equation}
An important alternative representation of $K(z)$ is given by
the integral ($\beta$, $z > 0$)
\begin{equation} \label{k2}
K(\sqrt{\beta} z) = \frac{1}{2 \pi i} \int_{-i \infty}^{i \infty} \!
dx \; \frac{e^{-i \beta x^2}}{x + z} \; .
\end{equation}
Due to this form the function $K$ arises in
semiclassical evaluations of oscillatory integrals in
which a pole of the integrand is close to a stationary point.
In figure \ref{figk} we show the real and imaginary parts of
the function $K(z)$ for positive $z$. The function has its
largest absolute value at $z=0$, and from approximately $z=3$
on it agrees well with its asymptotic approximation
(\ref{kasym}).

With the relation between the propagator and the Green 
function in (\ref{greenex}) one can obtain a uniform
approximation for the Green function. We consider
the contributions from the two parts of the propagator
in (\ref{propuni}) separately. These parts are called
the geometrical and the diffractive part in the following
\begin{equation} \label{greentot}
G_\alpha(\r,\ro,E) \approx G_g(\r,\ro,E) + G_d(\r,\ro,E) \; .
\end{equation}
For the first part one can evaluate the integral in 
(\ref{greenex}) by stationary phase evaluation and one
obtains, analogously to the propagator, the free
semiclassical Green function modified by a phase
\begin{equation} \label{greengeo}
G_g(\r,\ro,E) = \frac{M}{\hbar^2 \sqrt{2 \pi k |\r - \ro|}} \;
\exp \left\{ i k |\r - \ro| + i \alpha (\phi -\phi_0) 
- i \frac{3 \pi}{4} \right\} \; .
\end{equation}
For the second part we express the Fresnel function by the
integral (\ref{k2}) and arrive at
\begin{align} \label{laptra}
G_d(\r,\ro,E) & \approx 
- \frac{i M \sin(\alpha \pi)}{2 \pi^2 \hbar^2} 
\lim_{\varepsilon \rightarrow 0} \int_0^\infty \! dt 
\; \int_{-i \infty}^{i \infty} \! dz \;
\frac{1}{t} \, \frac{1}{z + \cos \frac{\phi - \phi}{2}}
\notag \\ & \qquad \qquad \times
\exp \left\{ \frac{iM}{2 \hbar t} (r + r_0)^2 +
\frac{i}{2} (\phi - \phi_0) 
- i \frac{2 M r r_0 z^2}{\hbar t} 
+\frac{i}{\hbar} t (E + i \varepsilon) \right\} \; .
\end{align}
This double integral has a stationary point at $(z,t)=(0,t_{cl})$
where $t_{cl} = \sqrt{M/(2E)}(r + r_0)$ is the classical time
for the path from $\r$ to $\r_0$ via the origin at energy $E$.
In order to evaluate the integral we expand the exponent
in (\ref{laptra}) up to second order around the stationary
point. Then the integral over $t$
can be evaluated by stationary phase approximation,
since the stationary point at $t=t_{cl}$ is well separated
from the pole at $t=0$. In the integral over $z$, however,
the $z$-dependence of the denominator has to be taken 
into account and this yields again a modified Fresnel function.
In this way a uniform approximation for the diffractive part
of the Green function is obtained
\begin{equation} \label{greenuni}
G_d(\r,\ro,E) = \frac{M \sin(\alpha \pi) \sqrt{2 \pi i}}{
\pi \hbar^2 \sqrt{k (r+r_0)}} \,
\exp \left\{ i k (r + r_0) + \frac{i}{2} (\phi -\phi_0) \right\}
\; K \left[ \sqrt{\frac{2 k r r_0}{r + r_0}} \, \cos
\left( \frac{\phi - \phi_0}{2} \right) \right] \; ,
\end{equation}
where again the angular coordinates have to be chosen such
that $|\phi - \phi_0| \leq \pi$. This approximation
for the Green function will be the ingredient for the
derivation of semiclassical contributions of diffractive
orbits in trace formulas in the next section. It is valid
as long as $k r$, $k r_0 \gg 1$.

We consider now an approximation to this expression that
is valid if $|\phi - \phi_0|$ is not too close to $\pi$.
Then the argument of the Fresnel function can be
replaced by its leading asymptotic term (\ref{kasym}) and
yields
\begin{equation} \label{greengtd}
G_d(\r,\ro,E) \sim
\frac{M \sin(\alpha \pi)}{2 \pi k \hbar^2 \sqrt{r r_0}
\cos \left( \frac{\phi - \phi_0}{2} \right) }
\exp \left\{ i k (r + r_0) + \frac{i}{2} (\phi - \phi_0)
+ i \frac{\pi}{2} \right\} 
\end{equation}
This approximation is of the general form that is
obtained within the Geometrical Theory of Diffraction (GTD)
(see e.\,g.\ \cite{Kel62}) 
\begin{equation} \label{gtd1}
G_d(\r,\ro,E) \approx \frac{\hbar^2}{2 M} \, G_0(\r,0,E) \,
{\cal D}(\phi,\phi_0) \, G_0(0,\ro,E) \; .
\end{equation}
In this theory the scattering is described by a
free Green function from $\r_0$ to the scattering 
source at the origin, multiplied by a diffraction
coefficient ${\cal D}(\phi,\phi_0)$ that contains
the information about the particular scattering process and a
further free Green function from the origin to the point $\r$.
A comparison with (\ref{greengtd}) shows that the diffraction
coefficient for the present case is given by
\begin{equation} \label{gtd2}
{\cal D}(\phi,\phi_0) = \frac{2 \sin(\alpha \pi)}{
\cos \left(\frac{\phi - \phi_0}{2} \right) }
\exp\left\{ i \frac{\phi - \phi_0}{2} \right\} \; .
\end{equation}
The term (\ref{greengtd}) can be interpreted as the contribution
of a trajectory that runs from $\r_0$ to the origin and then
to $\r$. It is of order $k^{-1/2}$ smaller than the contribution
of the direct trajectory from $\r_0$ to $\r$ in (\ref{greengeo}).
However, the GTD-approximation breaks down in the forward 
direction $|\phi-\phi_0|=\pi$ where the diffraction coefficient
(\ref{gtd2}) diverges. This reflects the fact that the diffractive
part of the Green function has a different leading asymptotic
term in the forward direction. Here the diffractive trajectory
contributes in the same order in $k$ as the direct trajectory.
The uniform approximation (\ref{greenuni}) interpolates between
these two different asymptotic regimes.

\begin{figure}[htb] \label{figgreen}
\begin{center}
\mbox{\epsfxsize8cm\epsfbox{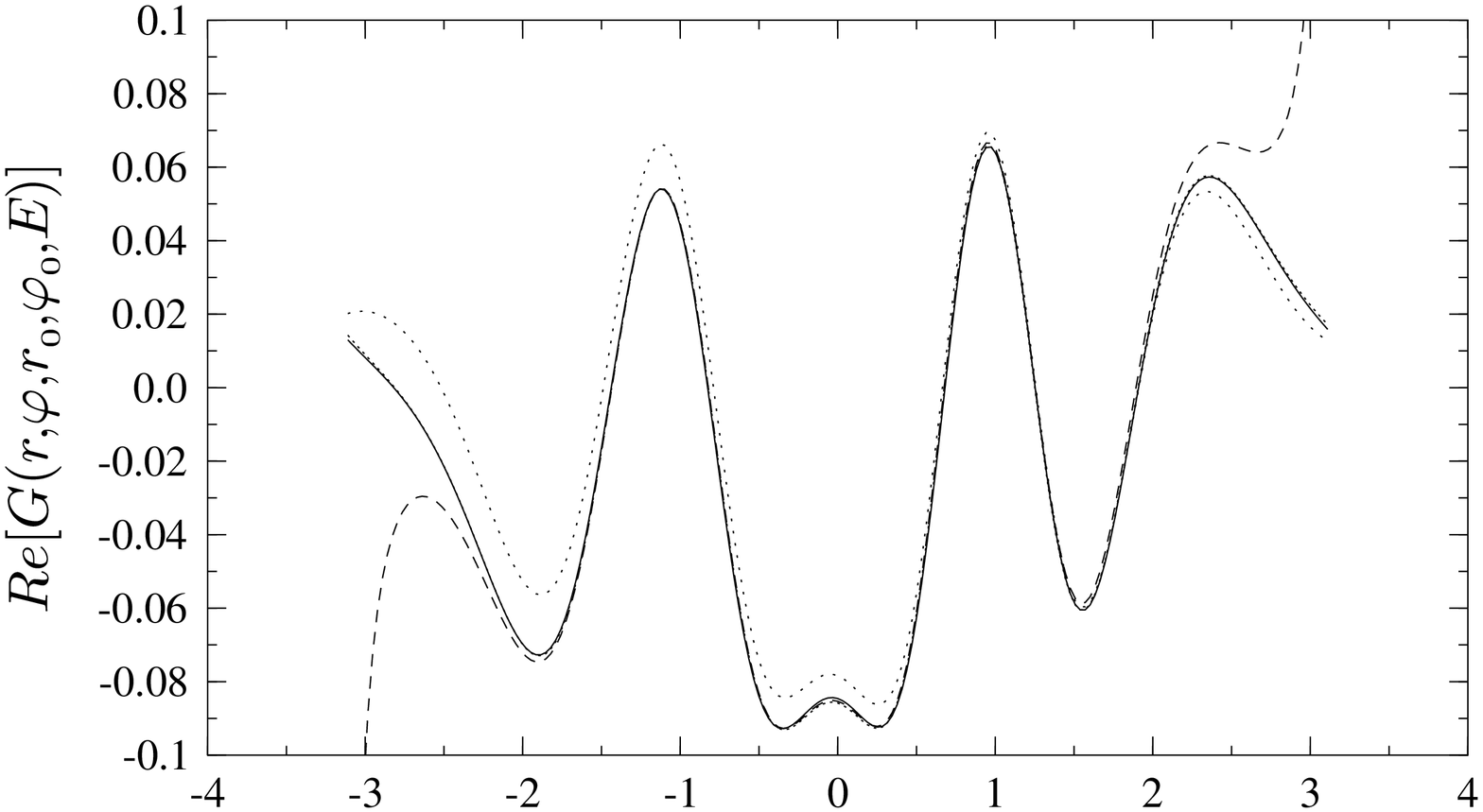}}
\mbox{\epsfxsize8cm\epsfbox{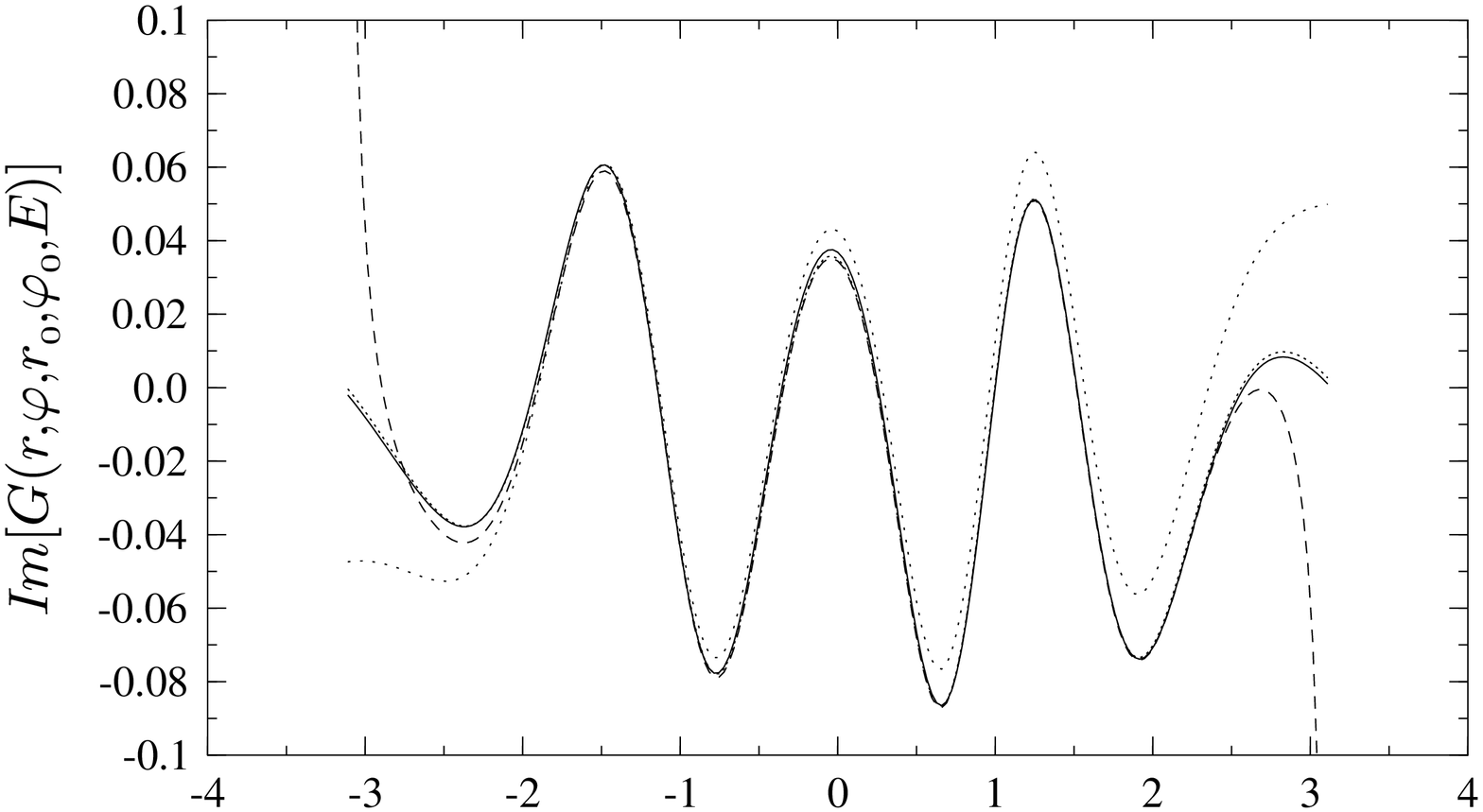}}
\end{center}
\caption{Real and imaginary parts for different approximations
of the Green function $G(\r,\r_0,E)$ with $r=1$, $r_0=2$,
$k=5$ and $\alpha = 0.4$. Full line: exact result, dotted line:
$G_g(\r,\r_0,E)$, dashed line: GTD-approximation,
and dashed-dotted line: uniform approximation.}
\end{figure}

In figure \ref{figgreen} we compare the different approximations
to the Green function with the exact result (\ref{greenex}).
The dotted line is the geometrical part (\ref{greengeo}).
The approximation is already reasonably good, but one
can still see a clear deviation from the exact curve.
By adding the diffractive part in the
GTD-approximation (dashed line) the difference becomes
much smaller for most values of $\phi - \phi_0$,
except near the endpoints $\phi - \phi_0 = \pm \pi$ where
it diverges. This divergence is removed by the uniform
approximation in (\ref{greenuni}) (dashed-dotted line)
for which the difference with the exact line can hardly
be seen, even though we chose relatively small values of $k r$
and $k r_0$.

\section{Semiclassical Contributions of Isolated Diffractive Orbits}
\label{secbound}

For the derivation of semiclassical approximations in billiard
systems it is convenient to apply the boundary integral method.
It provides an alternative formulation of the quantum mechanical
eigenvalue problem in terms of an integral equation along
the billiard boundary. All semiclassical contributions due to
diffraction in billiard systems that go beyond GTD have so far
been derived by this method \cite{PSSU96,SPS97}.

The boundary integral method has been developed for
billiard systems without internal fields (see e.g.\ 
\cite{KR74,Rid79}),
but it can be extended to more general situations. In \cite{TCA97}
Tiago et al.\ derived an integral equation for billiard
systems with magnetic field which is described by a vector
potential in the Coulomb gauge. The resulting integral
equation has the same form as for billiards without field.
For a billiard system with a flux line and Dirichlet boundary
conditions one obtains
\begin{equation} \label{inteq}
- 2 \int_{\partial {\cal D}} \! d s' \,
 \partial_{\hat{n}} \, G_\alpha(\r\,',\r,E) \, u(\r\,')
= u(\r)
\end{equation}
where from here on we use units in which $\hbar = 2M = 1$.
The difference to the field free case is that the Green function
in (\ref{inteq}) is the one for a flux line in a plane and
the solutions $u(\r)$ are the complex conjugate of the normal
derivatives of the wave functions. The form of the integral
equation in (\ref{inteq}) is different from that in
\cite{TCA97} since we applied an additional normal
derivative to the equation in order to obtain a non-divergent
integral kernel. Also the Green function differs by a factor
$(-8\pi)$ from that in \cite{TCA97}.

In (\ref{inteq}) the points $\r$ and $\r\,'$ lie on
the boundary, the position of $\r\,'$ is parametrised by $s'$,
and $\partial_{\hat{n}}$ denotes the outward normal derivative at
$\r$. The integral is evaluated along the boundary 
$\partial {\cal D}$ of the billiard system.
The left-hand-side of (\ref{inteq}) can be abbreviated
by $\hat{Q} u(\r)$ where $\hat{Q}$ is the integral operator
acting on $u$. Equation (\ref{inteq}) is a Fredholm equation of
the second kind and it has non-trivial solutions only if the determinant 
$\Delta(E) := \det(\boldsymbol{1} - \hat{Q}(E))$ vanishes.
This condition determines the quantum energies of the
problem. For a summary of the properties of the Fredholm
determinant $\Delta(E)$ for two-dimensional billiards without
fields with corresponding references see \cite{Sie98}.

From the condition of the vanishing of the Fredholm determinant
one can obtain an expression for the density of states in terms of
the integral operator $\hat{Q}$. It is given by
\begin{equation} \label{secc1}
d(k) = d_{\text{sm}}(k) + \frac{1}{\pi} \frac{d}{d k}  
\Im \sum_{n=1}^\infty \frac{1}{n} \Tr \hat{Q}^n (k)\; ,
\end{equation}
where
\begin{equation} \label{secc2}
\Tr \, \hat{Q}^n (k) = (-2)^n \int_{\partial {\cal D}}
\! d s_1 \dots d s_n \;
\partial_{\hat{n}_1} G_\alpha(\r_2,\r_1,E) \;
\partial_{\hat{n}_2} G_\alpha(\r_3,\r_2,E) \dots
\partial_{\hat{n}_n} G_\alpha(\r_1,\r_n,E) \; .
\end{equation}
The sum in (\ref{secc1}) is not convergent for real $k$,
and for the derivation we assume that $k$ has a sufficiently
large imaginary part and that $\Re k > 0$.
The density of states in (\ref{secc1}) has a smooth
part and a sum over oscillatory integrals. A semiclassical
evaluation of the oscillatory integrals yields the leading
semiclassical contributions to the density of states from
periodic orbits and diffractive orbits. These contributions
can be separated by writing $G_\alpha$ as sum of its geometrical
and its diffractive part. Then (\ref{secc2}) consists of
$2^n$ terms. The periodic orbit contributions are contained 
in the term that contains only geometrical Green functions.
By evaluating this term in stationary phase approximation
one obtains the usual Gutzwiller expression for isolated
periodic orbits \cite{Gut71}, modified by an additional phase
$2 \pi m \alpha$ where $m$ is the winding number of the orbit
around the flux line. 
The terms with $l$ diffractive Green functions on the other
hand contain the contributions of diffractive orbits that are
scattered $l$ times on the flux line.

In this article we consider only diffractive
orbits that are scattered once on a flux line. We thus 
have to evaluate an integral with a product of $n-1$ geometrical
Green functions and one diffractive Green function. 
This integral can be considerably simplified by applying a
composition law for the geometrical part of the Green functions.
\begin{equation} \label{secc5}
G_g^{(n)}(\r,\r\,',E) \approx
(-2)^n \int_{\partial {\cal D}}
\! d s_1 \dots d s_n \;
G_g(\r_1,\r\,',E) \; \partial_{\hat{n}_1}
G_g(\r_2,\r_1,E) \dots \partial_{\hat{n}_n}
G_g(\r,\r_n,E) \; .
\end{equation}
where the approximate sign denotes that the integrals
are evaluated in stationary phase approximation.
The composition law was proved for ordinary billiard systems
in \cite{SPS97} and it is a semiclassical version of the
multiple reflection expansion of the Green function of
Balian and Bloch. Since the presence of a flux line
adds only a phase to the Green function which doesn't
effect the stationary points in leading order, the
composition law holds also in the present case and 
the phase is simply additive. The semiclassical expression
for $G_g^{(n)}$ is then (see \cite{SPS97})
\begin{equation} \label{secc7}
G_g^{(n)} (\r_2,\r_1,E) = \sum_{\xi_n}
\frac{1}{\sqrt{8 \pi k |\tilde{M}_{12}|}}
\exp \left\{ i k \tilde{L} - i \frac{\pi}{2} \tilde{\nu}
- i \frac{3 \pi}{4} + i \alpha \phi_{21} \right\} \; .
\end{equation}
Here $\phi_{21}$ is the total winding angle of the 
trajectory around the flux line. It can be written
as $\phi_{21} = 2 \pi m + \Delta \phi$ where $m$ is
an integer and $|\Delta \phi| \leq \pi$. Furthermore,
$\tilde{M}_{12}$ is the $12$-element of the stability matrix,
$\tilde{L}$ is the length of the trajectory, and $\tilde{\nu}$
is the number of conjugate points from
$\r_1$ to $\r_2$. The stability matrix $\tilde{M}$ is evaluated
at unit energy and is energy independent. We use a tilde here in
order to distinguish the quantities from the corresponding 
ones for the diffractive orbit. With (\ref{secc5})
one obtains the following expression for the partial
contribution to the level density from diffractive orbits
with one scattering event
\begin{equation} \label{startap}
d_{part} = 
\frac{4}{\pi} \Im \frac{d}{d k} \int_{\partial {\cal D}}
\! d s_1 \, d s_2 \;
\partial_{\hat{n}_1} G_d(\r_2,\r_1,E) \;
\partial_{\hat{n}_2} G_g^{(n-1)}(\r_1,\r_2,E) \; .
\end{equation}

We insert now (\ref{secc7}) and (\ref{greenuni}) 
with (\ref{k2}) for the semiclassical and the
diffractive Green function, respectively. The normal derivative
gives in both cases in leading order a factor $i k \cos \alpha$
where $\alpha$ is the angle between the normal and the outgoing
trajectory. We obtain
\begin{equation}
d_{part}
= \Im \frac{d}{dk} \frac{k \cos \alpha_1 \cos \alpha_2
\sin(\alpha \pi)}{ 2 \pi^3 \sqrt{|\tilde{M}_{12}| (r_1 + r_2)}}
\exp \left\{ - i \frac{\pi}{2} \tilde{\nu} + i \alpha \phi_{21}
- i \frac{\Delta \phi}{2} \right\} \, I \; ,
\end{equation}
where
\begin{equation} \label{diffint}
I = \int_{\partial {\cal D}} \! d s_1 \, d s_2 \,
\int_{-i \infty}^{i \infty} \! dz \; 
\frac{\ds \exp \left\{ i k (\tilde{L} + r_1 + r_2) 
- i \frac{2 k r_1 r_2}{r_1 + r_2} z^2 \right\} }{\ds
z + \cos \frac{\Delta \phi}{2}} \; .
\end{equation}
The main part in the derivation consists in the evaluation of 
the diffraction integral $I$. If all three integrals are
evaluated by a stationary phase approximation
one obtains the contribution of the diffractive
orbit in the GTD approximation. This expression diverges
when $|\Delta \phi|=\pi$. In order to obtain a semiclassical
approximation that is uniformly valid for all angles one has
to take the dependence of the denominator on the integration
variables into account. One can do this by expanding the exponent
in (\ref{diffint}) up to second order in $s_1$, $s_2$ and $z$,
and the denominator up to first order in these variables.
In the denominator one obtains in this way
\begin{equation} \label{denom}
\cos \frac{\Delta \phi}{2} \approx
\cos \frac{\Delta \phi_0}{2} + \frac{1}{2}
\left( \frac{s_1 \cos \alpha_1}{r_1} - 
\frac{s_2 \cos \alpha_2}{r_2} \right)
\sin \frac{\Delta \phi_0}{2} \; .
\end{equation}
This method corresponds to a particular choice
of a uniform approximation. It yields an approximation
that is correct at the `optical boundary' $|\Delta \phi|=\pi$
and in the GTD limit,
and it interpolates between them. There are other possibilities
to obtain a uniform approximation, for example by mapping the
exponent onto a quadratic function in the three variables 
and then choosing an appropriate approximation for the
amplitude function. This reflects the fact that uniform
approximations are in general not unique. For example,
a different interpolating approximation can be obtained
by dropping the sine-term in (\ref{denom}) which corresponds
to replacing it by its value at the optical boundary.

The evaluation of the integral $I$ is rather lengthy. 
It consists in carrying out the expansions and performing
linear transformations of the variables such that, finally, the
denominator depends only on one of the variables and the integral
can be expressed in terms of the modified Fresnel function.
A comparison with \cite{SPS97} shows, however, that the same
diffraction integral occurs for the diffraction on billiard
corners. For that reason, the calculations don't have to be
done again and the result can be inferred from this paper.
One has ($c>0$)
\begin{align} \label{intres}
& \int_{-\infty}^\infty \! d s_1 \! 
  \int_{-\infty}^\infty \! d s_2 \! 
  \int_{-i \infty}^{i \infty} \! dz \;
\frac{\ds \exp \left\{ i k (\tilde{L} + r_1 + r_2 - c\,z^2) \right\}
}{\ds a + z + b \left( \frac{s_1 \cos \alpha_1}{2 r_1}
               - \frac{s_2 \cos \alpha_2}{2 r_2} \right) }
\notag \\[5pt] = &
- \frac{4 \pi^2 \sign(a) \tau}{k |b| \cos \alpha_1
\cos \alpha_2} \sqrt{\frac{2 r_1 r_2 |\tilde{M}_{12}|}{c |\Tr M - 2|}}
\, K \left[ i^{\ds \kappa} \left| \frac{a}{b} \right|
\sqrt{\frac{2 k |M_{12}|}{|\Tr M - 2|}} \right]
\text{\Large e}^{\ds i k L - i \pi(\nu + \kappa - \tilde{\nu})/2}
+ b.c. \; ,
\end{align}
where b.c.\ denotes a boundary contribution whose origin is the
discontinuity of the diffractive part of the Green function.
It is cancelled by a corresponding boundary contribution from
the geometrical part of the Green function. In (\ref{intres}) 
$L$ and $M$ denote the length and stability matrix of the
orbit, respectively, and $\nu$ is the number of conjugate
points along the orbit. $\kappa$ is given by
\begin{equation}
\kappa = \begin{cases}
0 & \text{if} \frac{M_{12}}{\Tr M - 2} > 0 \, , \\
1 & \text{if} \frac{M_{12}}{\Tr M - 2} < 0 \, ,
\end{cases}
\end{equation}
and $\tau = 1 - 2 \kappa$.

With (\ref{intres}) and noting that the derivative gives in
leading order a factor $(i L)$ one obtains the final result
for the contribution of a diffractive orbit $\xi$
\begin{equation} \label{final}
d_\xi(k) = - \Re \left( \frac{\ds 2 \tau L \sin(\alpha \pi) \,
\text{\Large e}^{\ds i k L - i \pi \mu /2
+ i \phi_{21} \alpha - i \Delta \phi/2} }{\ds \pi \,
|\sin(\frac{\Delta \phi}{2})| \, \sqrt{|\Tr M - 2|}}
\, K \left[ i^{\ds \kappa} 
\left| \cot \frac{\Delta \phi}{2} \right|
\sqrt{\frac{2 k |M_{12}|}{|\Tr M - 2|}} \right] \,
 \right)
\end{equation}
where $\mu = \nu + \kappa$ agrees with the usual
definition of the Maslov index for periodic orbits.

Formula (\ref{final}) 
is the main result of this article. It describes the
contribution of a diffractive orbit to the density of states
and is valid for all scattering angles $\Delta \phi$, and
for $0 < \alpha < 1$. A slightly simpler approximation can
be obtained by dropping the sine-term in (\ref{final}) and
replacing the cotangent by a cosine, corresponding to the
discussion after (\ref{denom}).

The use of the full Fresnel function in (ref{final})
is necessary in an angular region around the forward 
direction with width of order $k^{-1/2}$.
For angles outside this range the Fresnel function can be
replaced by its leading asymptotic form which results in
\begin{equation} \label{finlim}
d_\xi(k) \approx \Re \left( \frac{L \sin(\alpha \pi)}{\pi
\cos(\Delta \phi/2)} \frac{1}{\sqrt{2 \pi k |M_{12}|}}
\exp \left\{ i k L - i \frac{\pi}{2} \nu
+ i \alpha \phi_{21} - \frac{i}{2} \Delta \phi 
- i \frac{3 \pi}{4} \right\} \right) \; .
\end{equation}
This agrees with the GTD approximation for the diffractive orbit
\cite{GTD}
\begin{equation} \label{fingtd}
d_\xi(k) = \Re \left( \frac{L}{\pi} G^\xi_g(\r_2,\r_1)
{\cal D}(\phi_1,\phi_2) \right) \, .
\end{equation}
It is a contribution that is of order $\sqrt{k}$ smaller
than the contribution of a isolated periodic orbit.
In (\ref{fingtd}) $G_g^\xi$ denotes the contribution 
from the trajectory $\xi$ to the geometrical part of
the Green function where $\r_1=\r_2$ is the position
of the flux line.

\begin{figure}[htb] \label{figtraj}
\begin{center}
\mbox{\epsfxsize8cm\epsfbox{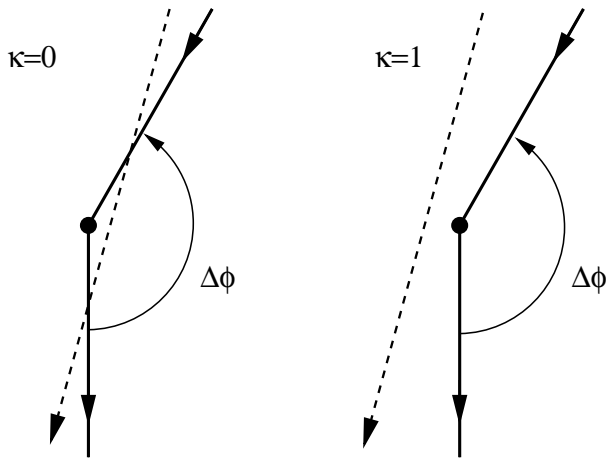}}
\end{center}
\caption{Parts of a diffractive orbit and a nearby periodic
orbit for $\kappa=0$ and $\kappa=1$.}
\end{figure}

In the opposite limiting case at $|\Delta \phi| = \pi$
the uniform approximation is of the same order as
that of a periodic orbit and it is discontinuous. As $|\phi|$
goes through $\pi$ the contribution makes a step of size
\begin{equation} \label{discon}
\left. d_\xi(k) \right|_{\Delta \phi = \pi-0} - 
\left. d_\xi(k) \right|_{\Delta \phi = -\pi+0} = 
- \frac{2 \tau L \sin(\alpha \pi)}{\pi \sqrt{|\Tr M - 2|}}
\sin \left(k L - \frac{\pi}{2} \mu + (2m+1) \pi \alpha \right) \, .
\end{equation}
In order to understand this discontinuity we consider
an angle $\Delta \phi = \pi - \varepsilon$ which is 
infinitesimally different from $\pi$. By examining the
linearised motion around the diffractive orbit one
finds that there is, in general, a periodic orbit in an
infinitesimal neighbourhood of the diffractive orbit. This orbit
is obtained from the conditions that $dq=dq'$ and $dp=dp'-k\varepsilon$
where the primed and the unprimed quantities are infinitesimal
perpendicular deviations from the starting point and
endpoint of the diffractive orbit, respectively. From these
conditions one finds that the position of the periodic orbit
is given by $dq=-k \varepsilon M_{12}/(\Tr M - 2)$.
Depending on the sign of the left-hand-side, i.e.\ on the value
of $\kappa$, there can be two possible cases that are shown in
figure \ref{figtraj} for positive $\Delta \phi$ near $\pi$. 

If one moves the flux line in a way that $\Delta \phi$ goes 
through $\pi$ then the flux line crosses this periodic
orbit at the same instant. For the periodic orbit the
semiclassical contribution is also discontinuous since
the winding number of the periodic orbit changes. For
a primitive orbit the winding number changes by $+1$
if $\kappa=0$ and by $-1$ if $\kappa=1$. By writing
the product of sines in (\ref{discon}) as a sum of
two cosines one can show that the two discontinuities
cancel exactly and the sum of both contributions is
continuous. This shows that the uniform approximation
for diffractive contributions is necessary in order
to make semiclassical approximations continuous
if the position of a flux line is changed.

\section{Diffractive orbits in the circular billiard}
\label{seccirc}

A simple example in which diffractive orbits are not
isolated is a circular billiard with a flux line in
its centre. This system is integrable since the energy
and the angular momentum around the centre are
conserved. All diffractive orbits run from the flux 
line to the boundary and are reflected back directly
onto the flux line. They appear in one- or more-parameter
families, and their lengths are multiples of $2R$ where
$R$ is the radius of the circle. Contributions of these
diffractive orbits to the density of states have been observed
in \cite{RBMBM96}.

We are interested in the leading order semiclassical
contributions of the diffractive orbits. For this
purpose one cannot apply formula (\ref{final}) since
the orbits have a stability matrix with trace two.
Although one can in principle obtain the diffractive
contributions from the boundary element method, it is
now much more convenient to start from the
torus-quantisation conditions for integrable systems
(EBK-conditions). 
In this way one can obtain all leading order diffractive
contributions to the density of states at once, even those
for multiple diffraction.

The exact solutions of the Schr\"odinger equation are
given by the solutions of the flux line in a plane
(\ref{solpla}) with a different normalisation constant
and the additional condition that the wave functions
have to vanish at $r=R$. From this condition it 
follows that the energies are determined by the
zeros of the Bessel functions
$E_{m,n} = \hbar^2 k_{m,n}^2/(2M)$, where
$k_{m,n} = j_{|m-\alpha|,n+1}/R$, and $j_{\nu,n}$ is the
$n$-th zero of the Bessel function with index $\nu$.

The EBK-quantisation conditions yield semiclassical 
approximations to these energy levels. They have the form 
\begin{equation} \label{ebk1}
\oint \! d\phi \; p_\phi = 2 \pi \hbar m \; , \; \; 
m \in \mathbb{Z} \; , \qquad \quad
\oint \! dr \; p_r = 2 \pi \hbar \left( n + \frac{3}{4} \right) \; ,
\; \;  n = 0, 1, 2, \dots
\end{equation}
Due to the conservation of the angular momentum the first
condition gives $p_\phi = \hbar m$ and the second
condition is evaluated with the energy conservation law
$ E = ( p_r^2 + (p_\phi - \hbar \alpha)^2/r^2)/(2M)$
and yields 
\begin{equation} \label{ebk2}
\sqrt{k_{m,n}^2 R^2 - (m-\alpha)^2} - |m-\alpha| \, \arccos
\frac{|m-\alpha|}{k_{m,n} R} = \pi \left( n + \frac{3}{4} \right)
\end{equation}
which determines $k_{m,n}$ as a function of the two quantum
number $m$ and $n$.

From (\ref{ebk2}) the periodic orbit contributions to the
density of states are obtained by applying the Poisson summation
formula \cite{BT76}
\begin{align} \label{poisson}
d(k) & =  \sum_{M,N=-\infty}^\infty
\int_{-\infty}^\infty \! dm \; \int_{-3/4}^\infty \! dn \;
e^{2 \pi i (Mm + Nn)} \, \delta ( k - k_{m,n}) 
\notag \\ &
=  \sum_{M,N=-\infty}^\infty
\int_{\alpha-kR}^{\alpha+kR} \! dm \;
\frac{1}{\pi k} \sqrt{k^2 R^2 - (m - \alpha)^2} \,
\exp\{2 \pi i (Mm + Nn(m,k))\}
\notag \\ &
=  \sum_{M,N=-\infty}^\infty
\int_0^{kR} \! dm \;
\frac{1}{\pi k} \sqrt{k^2 R^2 - m^2} \times
\notag \\ & \qquad
\exp \left\{ 2 \pi i M (m+\alpha) + 2 i N \left[
\sqrt{k^2 R^2 - m^2} - m \arccos \frac{m}{kR} \right] 
- \frac{3 \pi}{2} i N \right\}
+ [\alpha \rightarrow -\alpha] \; .
\end{align}

We consider now the main semiclassical contributions to the
integrals in (\ref{poisson}). Altogether these are four
contribution. One is the only non-oscillatory term $M=N=0$,
which yields the leading area-term for the mean density of states 
\begin{equation}
d_A(k) = \frac{2}{\pi k} \int_0^{kR} \! dm \; \sqrt{k^2 R^2 - m^2} 
= \frac{1}{2} k R^2 \; .
\end{equation}
This agrees with the leading term in Weyl's law $Ak/(2 \pi)$
with area $A=\pi R^2$.

The second main contributions arises from the stationary points
of the integrals. A stationary phase approximation gives the
contributions of the periodic orbits that have been obtained
also in \cite{RBMBM96}.
\begin{equation} \label{circpo}
d_{po}(k) = \sum_{N=2}^\infty \sum_{M+1}^{[N/2]} g_{M,N}
\sqrt{\frac{4k}{\pi N} R^3 \sin^3 \frac{\pi M}{N} }
\cos \left( 2 N k R \sin \frac{\pi M}{N} - \frac{3 \pi}{2} N 
+ \frac{\pi}{4} \right) \, \cos (2 \pi M \alpha)
\end{equation}
where $g_{M,N} = 1$ if $M  =   N/2$
and   $g_{M,N} = 2$ if $M \neq N/2$.
This factor arises from the fact that the stationary points
for $M=N/2$ lie on the endpoint $m=0$ of the integrals in
(\ref{poisson}) and give only half the contribution.

The remaining two contributions follow from the boundaries
of the oscillatory integrals. They can be obtained by an
integration by part which yields
\begin{equation} \label{bc}
\text{b.\,c.\ of} \; \left\{
\int_{-\infty}^z \! dx \; g(x) \, e^{if(x)} \right\}
= - i \frac{g(z)}{f'(z)} \, e^{if(z)} \; ,
\end{equation}
for an upper end of an integration range. If $z$ is on the
lower end of an integration range the end point contribution
is given by minus the right hand side of (\ref{bc}).

With (\ref{bc}) one obtains for the contribution from $m=kr$
\begin{align} \label{peri}
d_L(k) & =  \sum_{N,M=-\infty}^\infty \!\!\!\!\!\!{}' \, \, 
\frac{1}{\pi k} \left. \frac{\sqrt{k^2 R^2 - m^2}
\exp \left\{ 2 \pi i M (m+\alpha) - \frac{3 \pi}{2} i N \right\} }{
2 \pi i M - 2 i N \arccos \frac{m}{k R} } \right|_{
m \rightarrow k R} + [\alpha \rightarrow -\alpha] 
\notag \\ &
= \frac{2 R}{\pi} \sum_{N=1}^\infty \frac{\sin \left(
\frac{3 \pi}{2} N \right)}{N}
\notag \\ & = - \frac{R}{2} \; .
\end{align}
where the prime denotes that the term $(N,M)=(0,0)$ is excluded
from the sum since it corresponds to a non-oscillatory integral.
The result in (\ref{peri}) can be identified with the perimeter
term in the asymptotic expansion of the mean density of states.

From the other end points of the integrals at $m=0$ one obtains
\begin{align}
d_d(k) & = \frac{iR}{2 \pi^2}
\sum_{\substack{M,N=-\infty \\ M \neq N/2}}^\infty
\frac{1}{M - \frac{N}{2}} \exp \left\{
2 \pi i M \alpha + 2 i N k R - \frac{3 \pi}{2} i N \right\}
+ [\alpha \rightarrow -\alpha] 
\notag \\ & = d_d^e(k) + d_d^o(k) \; .
\end{align}
Here the terms $M=N/2$ have to be excluded from the sum since
they have already been taken into account by the stationary
phase evaluation. For convenience the sum is split into
two parts, corresponding to a summation over even and odd
values of $N$, respectively.

For the first part $N$ is replaced by $2N$, the summation index $M$
is shifted by $N$, and terms for positive and negative values
of $M$ are combined 
\begin{align} \label{de}
d_d^e(k) & =- \frac{R}{\pi^2}
\sum_{N=-\infty}^\infty \sum_{M=1}^\infty
\frac{\sin [ 2 \pi \alpha M]}{M} \exp \left\{
i 2 N (\pi \alpha + 2 k R - \frac{3 \pi}{2})\right\}
+ [\alpha \rightarrow -\alpha] 
\notag \\ & 
= \frac{2 R}{\pi} (1 - 2 \alpha) \sum_{N=1}^\infty  
\sin \left( 2N(2kR - \frac{3 \pi}{2}) \right)
\sin(\pi \alpha 2N) \; ,
\end{align}
and for the second part $N$ is replaced by $2N+1$, the summation
index $M$ is shifted by $N$, and terms for $M \geq 1$ and $M<1$
are combined
\begin{align} \label{do}
d_d^o(k)
 & = - \frac{2 R}{\pi^2}
\sum_{N=-\infty}^\infty \sum_{M=1}^\infty
\frac{\sin [ \pi \alpha ( 2 M - 1) ]}{2 M - 1} \exp \left\{
i (2 N + 1)(\pi \alpha + 2 k R - \frac{3 \pi}{2})\right\}
+ [\alpha \rightarrow -\alpha] 
\notag \\ & 
= \frac{2 R}{\pi} \sum_{N=0}^\infty  
\sin \left( (2N+1)(2kR - \frac{3 \pi}{2}) \right)
\sin(\pi \alpha (2N+1)) \; .
\end{align}
The formulas for evaluating the summations in (\ref{peri}),
(\ref{de}) and (\ref{do}) can be found in \cite{GR80} and
are valid for $0 < \alpha < 1$.

Collecting all results, the complete trace formula is given by
\begin{align} \label{trace}
d(k) \approx \; & \bar{d}(k) 
+ \sum_{N=2}^\infty \sum_{M=1}^{[N/2]} g_{M,N}
\sqrt{\frac{4k}{\pi N} R^3 \sin^3 \frac{\pi M}{N} }
\cos \left( 2 N k R \sin \frac{\pi M}{N} - \frac{3 \pi}{2} N 
+ \frac{\pi}{4} \right) \, \cos (2 \pi M \alpha)
\notag \\ & 
+ \sum_{N=1}^\infty \frac{2R}{\pi} \sin \left( 
2 N k R - \frac{3 \pi}{2} N \right) \sin(\pi N \alpha)
- 2 \alpha \sum_{N=1}^\infty \frac{2R}{\pi} \sin \left( 
4 N k R - 3 \pi N \right) \sin(2 \pi N \alpha) \; .
\end{align}
It consists of the mean level density, the contributions
of periodic orbits, and the contributions of diffractive orbits
which are by an order $k^{-1/2}$ smaller than those of the
periodic orbits. The formula is similar to the corresponding
one for a harmonic oscillator with a flux line \cite{BBLMM95}.

The diffractive orbit term in (\ref{trace}) can be given a more
direct interpretation since it can be transformed into a sum
of delta functions by using the Poisson summation formula.
One obtains
\begin{align} \label{ddiff}
d_d(k) = & - \frac{R}{2 \pi} \sum_{n=-\infty}^\infty
\left[ \delta \left( \frac{kR}{\pi} - \frac{3}{4} +
                     \frac{\alpha}{2} + n \right) 
-      \delta \left( \frac{kR}{\pi} - \frac{3}{4} -
                     \frac{\alpha}{2} + n \right) \right]
\notag \\ & + \frac{\alpha R}{\pi} \sum_{n=-\infty}^\infty
\left[ \delta \left( \frac{2kR}{\pi} - \frac{3}{2} +
                     \alpha + n \right) 
-      \delta \left( \frac{2kR}{\pi} - \frac{3}{2} -
                     \alpha + n \right) \right]
\end{align}
The argument of one of the delta functions can be
identified with the semiclassical quantisation 
condition for eigenvalues with vanishing angular
momentum ($m=0$): $kR \approx \pi \alpha / 2 
+ 3 \pi/4 + n \pi$. Since the full trace formula
produces delta peaks at the eigenvalues given by
the EBK conditions (\ref{ebk2}), the diffractive
orbits have the following role: they contribute
to peaks at eigenvalues with vanishing 
angular momentum and they cancel wrong peaks
that are produced by the periodic orbit sum.

The fact that the periodic orbits produce wrong
peaks can be understood by noting that the periodic
orbit contributions (\ref{circpo}) are invariant
under $\alpha \rightarrow -\alpha$. Due to this
symmetry the periodic orbits give also rise to
wrong peaks at $kR \approx - \pi \alpha / 2 
+ 3 \pi/4 + n \pi$. The first two terms in
(\ref{ddiff}) provide a correction to this failure
in the approximation for states with zero angular
momentum. The other two terms in (\ref{ddiff}) are
necessary in order that the approximation is invariant
under $\alpha \rightarrow 1 - \alpha$, since the spectrum
is invariant under this replacement.

\begin{figure}[htb] \label{figcirc}
\begin{center}
\mbox{\epsfxsize8cm\epsfbox{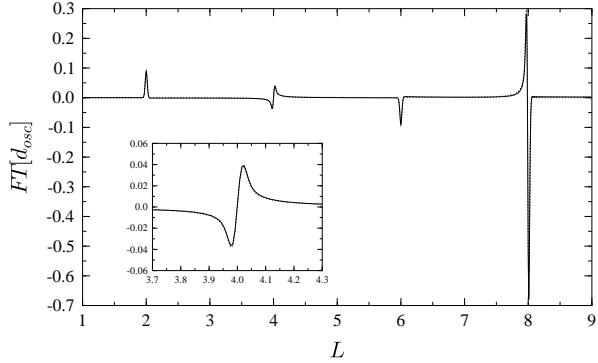}}
\end{center}
\caption{Fourier transform of the oscillatory part of the
density of states for the circular billiard with $\alpha=0.25$
and $R=1$ (full line) in comparison with the semiclassical
approximation (dotted line). Inset: Magnification of one peak.}
\end{figure}

A test of the trace formula (\ref{trace}) is presented
in figure \ref{figcirc} which shows a comparison
between quantum mechanical and semiclassical results. The 
value of $\alpha$ is chosen to be $\alpha=0.25$, since
for this value the periodic orbit contributions start
at $L=8R$. The difference between the two curves cannot
be seen on this scale. 

\section{Conclusions}
\label{seccon}

One general property of uniform approximations for
isolated diffractive orbits is that they make
semiclassical approximations continuous as a system
parameter is changed. They cancel discontinuities in
semiclassical contributions of periodic orbits.
In billiards with concave boundaries or with corners
those discontinuities are connected with the 
appearance or disappearance of periodic orbits.
In the present case the discontinuity is due
to a phase change in periodic orbit contributions.
Besides that, the diffraction on a flux line is
very similar to the diffraction on corners. The
final formula is expressed in terms of the 
same interpolating function, but it is also 
simpler since it consists only of one term
instead of four. This suggests, for example,
that a semiclassical study of two-dimensional
systems which are non-integrable but in which the
classical motion is still restricted to a two-dimensional
surface in phase space might be performed more easily
on billiards with a flux line instead of pseudo-integrable
polygonal billiards.

In this article we considered only diffractive orbits
that are scattered once on a flux line. The same method
can be used to obtain semiclassical contributions of
diffractive orbits with multiple scattering, but the
formulas become increasingly more complex and have to be
expressed in terms of multiple Fresnel integrals.
The treatment of an integrable billiard with a flux
line like the circular billiard is much simpler. There
all leading order diffractive contributions including
those from multiple diffraction can be obtained in one
step from the EBK conditions. Moreover, they can be
summed up and shown to contribute only to states with zero
angular momentum.

One remaining question is whether in semiclassical
arguments concerning spectral statistics in chaotic systems
it is sufficient to consider only periodic
orbits. One argument for this is that the semiclassical
contributions of most diffractive orbits (GTD region)
are by an order $1/\sqrt{k}$ smaller than those of
periodic orbits where $k$ is the wavenumber. 
For scattering angles near the forward direction
they contribute more strongly, up to the order of
a periodic orbit, but the corresponding angular regime
of this transitional region decreases proportional
to $1/\sqrt{k}$. So one might argue that
diffractive orbits become less and less important in
the semiclassical regime. 

Let us discuss this point in more detail. By applying
the trace formula in order to resolve adjacent energy
levels with wave numbers of order $k$ one has to take
into account all orbits up to the Heisenberg length
$L_H \propto k$. In a chaotic system the number of
diffractive orbits increases exponentially with 
the orbit length. If we assume that the scattering
angle is uniformly distributed then
the relative number of diffractive orbits in the
transitional region decreases like $1/\sqrt{k}$.
However, the total number of diffractive orbits
in the transitional region is still increasing
exponentially due to the exponential increase 
of all diffractive orbits. A similar
argument can be applied to orbits with multiple
scattering. It is not obvious that these orbits
can be neglected. Moreover one can show that 
even in the GTD approximation diffractive orbits
have a non-vanishing influence on spectral statistics
in the semiclassical limit \cite{Sie99}.
\bigskip \bigskip \bigskip

\noindent
{\large \bf Acknowledgements}
\bigskip

\noindent
Financial support by the Deutsche Forschungsgemeinschaft
in form of a `Habilitandenstipendium' (SI 380/2-1) 
is gratefully acknowledged.

\end{document}